# Bulk Superconductivity below 6K in PdBi$_2$Te$_3$ topological single crystal

M. M. Sharma[1,2], Lina Sang[3], Poonam Rani[1], X. L. Wang[3] and V.P.S. Awana[1,2,*]

[1]*CSIR-National Physical Laboratory, K.S. Krishnan Marg, New Delhi-110012, India*
[2]*Academy of Scientific and Innovative Research (AcSIR), Ghaziabad - 201002, India*
[3]*Institute of superconducting and electronic materials, University of Wollongong, NSW 2522, Australia*

## ABSTRACT

We study the structural and bulk superconducting properties of self flux grown PdBi$_2$Te$_3$ single crystal. Phase purity of as grown crystal is confirmed by Rietveld refinement of gently crushed powder XRD of the same. PdBi$_2$Te$_3$ crystallizes in rhombohedral structure with R-3 m space group along with small impurity of Bi. Scanning Electron Microscopy (SEM) images showed layered structure and the elemental analysis by energy dispersive X Ray analysis (EDAX) done on same confirmed the stoichiometry to be near to PdBi$_2$Te$_3$. Characteristic vibrational modes viz. A$^1_{1g}$, E$g^2$, A$^2_{1g}$ are clearly observed in Raman spectrum, and are slightly shifted from that as in case of Bi$_2$Te$_3$. Bulk superconductivity is confirmed by FC and ZFC magnetization measurements (M-T) exhibiting diamagnetic transition with T$_c^{onset}$ at around 6K. M-H plots at different temperatures of 2K, 2.5K, 3K, 3.5K, 4K, 4.5K, 5K and 6K showed clear opening of the loop right up to 6K. Both M-T and M-H clearly establish the appearance of bulk type II superconductivity below 6K in studied PdBi$_2$Te$_3$. The lower critical field H$_{c1}$ and upper critical field H$_{c2}$ are at 180Oe at 4800Oe respectively at 2K for as grown PdBi$_2$Te$_3$ crystal. Other critical parameters of superconductivity such as coherence length, penetration depth and kappa parameter are also calculated.

Key Words: Doped Topological Insulators, Crystal Growth, Structural Details, and Superconductivity

PACS: 74.70.Dd, 74.62.Fj, 74.25.F-

[*]**Corresponding Author**

Dr. V. P. S. Awana: E-mail: awana@nplindia.org
Ph. +91-11-45609357, Fax-+91-11-45609310
*Homepage: awanavps.webs.com*

## INTRODUCTION

Presently, condensed matter scientists are giving a lot of attention towards the topological properties of materials. Topological insulators (TIs) exhibit exotic quantum phenomenon due to their topologically protected conducting surface states associated with insulating gap in bulk [1-3]. It has been well established that superconductivity can be induced in TIs by proper doping and making a new class of materials called topological superconductors (TS) [4-12]. A topological superconductor is a fully gapped superconductor with odd parity pairing symmetry having conducting surface/edge states [13]. These materials are the topic of great interest in recent days due to their potential use in fault tolerant high speed quantum computing devices [14]. These materials also have fundamental importance due possibility of hosting most sought particles known as Majorana fermions [15]. Topological superconductors also show many novel properties such as delocalized Andreev Bound states (ABS) [16] and zero bias conduction peak [17] etc. These facts make it important to search for materials that can show topological superconductivity (TS). There are several studies that have shown the superconductivity in $Bi_2Se_3$ by doping of Copper (Cu) [4, 5], Strontium (Sr) [6, 7] and Niobium (Nb) [8, 9]. Somehow topological superconductivity in doped $Bi_2Te_3$ is not studied to that extent as for $Bi_2Se_3$. $Bi_2Te_3$ can be made superconducting by doping of Thallium (Tl) [10] and Palladium (Pd) [11, 12].

In this study we report crystal growth, structural and magnetic characterization of $PdBi_2Te_3$ to explore the superconductivity in $Bi_2Te_3$. It is important to study this material as there are only two reports available in literature on superconducting properties of $PdBi_2Te_3$ [11, 12]. We found here bulk superconductivity at below 6K in magnetization versus temperature (M-T) measurements of $PdBi_2Te_3$. M-H curves at different temperatures shows the presence of type-II superconductivity in studied sample with lower critical field at 180Oe and upper critical field of 4800Oe at 2K.

## EXPERIMENTAL

$PdBi_2Te_3$ crystal is synthesized by solid state reaction route via vacuum encapsulation. High quality 5N powders of Palladium (Pd), Bismuth (Bi), and Tellurium (Te) were taken in stoichiometric ratio to obtain 1 gram powder of $PdBi_2Te_3$, then it was ground using agate mortar till 30 minutes to get a homogenous powder inside Ar filled glove box. The grounded sample was palletized by using hydraulic pressure palletizer in rectangular shape. Then this pallet was put into a quartz tube and vacuum encapsulated up to a pressure of $2\times10^{-5}$ mbar, which was subsequently put into programmable muffle furnace and heated to $850^{o}C$ at a rate of $2^{o}C/min$. This sample was put on hold at $850^{o}C$ for 24 hours and was slowly cooled to $570^{o}C$ at a rate of $1^{o}C/hour$ and then annealed at same temperature for 24 hours. Finally, the sample is quenched in ice water to avoid the formation of any low temperature impurity phase viz. α-BiPd [18]. In another article published recently, the sample is slow cooled to room temperature from $650^{0}C$, which resulted in formation of some impurity phases along with main $PdBi_2Te_3$ [12]. In order to avoid the formation of un-reacted phases and segregation of Pd, we quenched the sample from $570^{o}C$. The situation is similar as we observed recently in case of $Nb_{0.25}Bi_2Se_3$ [19], where it is necessary to quench to sample from high temperature to avoid Nb segregation. After quenching, in ice water a silvery shiny and easily cleavable crystal of $PdBi_2Te_3$ is obtained. A schematic of followed heat treatment is shown in Fig. 1. Image of thus obtained crystal is shown in inset of Fig. 1. Rigaku made Mini Flex II X-ray diffractometer having CuKα radiation of 1.5418Å wavelength used for X Ray analysis of powder of as grown $PdBi_2Te_3$ crystal. Morphology and elemental analysis of as grown crystal is done by using Scanning Electron Microscope (SEM) and Energy Dispersive Spectroscopy (EDS) made from Bruker. Renishaw Raman Spectrometer is used to record Raman spectra of as grown crystal. QD-PPMS (Physical Property Measurement System) is used to do magnetic study of the as grown crystal of $PdBi_2Te_3$.

**RESULTS AND DISCUSSION:**

Rietveld refinement of powder XRD pattern of gently crushed as grown $PdBi_2Te_3$ crystal done by using Full Proof software is shown in Fig. 2. Besides the main $PdBi_2Te_3$ phase, some impurity of monoclinic Bi is also seen. The parameter of goodness of fit i.e. $\chi^2$ is found to be 2.85, which is in an acceptable range confirming that the crystal is grown mainly in $Bi_2Te_3$ phase having rhombohedral structure with R-3 m space group. Unit cell of as grown $PdBi_2Te_3$ crystal is constructed by using Vista software and is shown in inset of Fig. 2. The quintuple layers of $Bi_2Te_3$ are separated by van der waals gap. This van der waals gap provides opportunity of intercalation of atoms in $Bi_2Te_3$ and it is assumed that Pd atoms are incorporated in these gaps in $Bi_2Te_3$ unit cell. However, the presence of monoclinic Bi impurity, although in much less quantity than the main phase, calls for the Bi vacancies or Bi-site Pd substation in PdBi$_2$Te$_3$ main phase. Worth mentioning is the fact the only two reports [11, 12] yet available on PdBi$_2$Te$_3$ topological superconductor (TS) have also reported presence of other phases. Interestingly enough, though the impurity phases are mentioned in main text of ref. 11, the XRD is not shown, hence it is difficult to comment on quality of studied crystal and to compare. In other report i.e., ref. 12, the rietveld refinement of the PXRD of crystal is shown having various substantial impurity phases. The possible reason seems to be the slow cooling process to room temperature being used in ref. 11. Although not completely free from impurity, yet one can safely conclude that the quality of presently studied PdBi$_2$Te$_3$ is better than earlier available reports [11, 12]. Atomic positions of elements are found to be Bi(0,0,0.4039), Te(1) (0,0,0), Te(2) (0,0,0.2039).Lattice parameters thus obtained are as follows a=b= 4.4213Å & c= 30.3371Å, α = β = 90° & γ = 120°. The c parameter of PdBi$_2$Te$_3$ is found to slightly decrease to 30.337Å from 30.385Å of pure Bi$_2$Te$_3$. It is hard to tell whether intercalation increases or decreases c parameter, there are reports on both increment and decrement of c parameter by intercalation [20, 21], in our case c parameter is slightly decreased by intercalation. Further the small impurity of Bi may call for possible Bi (larger) site Pd (smaller) substitution in main PdBi$_2$Te$_3$.

Morphology of as grown PdBi$_2$Te$_3$ crystal is viewed by scanning electron microscopy. SEM image of as grown PdBi$_2$Te$_3$ crystal with 2μm resolution is shown in Fig. 3(a). SEM image shows a smooth layered structure confirming the laminar growth of grown PdBi$_2$Te$_3$ crystal. Elemental analysis of as grown PdBi$_2$Te$_3$ crystal is done by EDAX. EDAX spectrum of as grownPdBi$_2$Te$_3$ is shown in Fig. 3(b). EDAX spectrum confirms the presence of all elements in close to desired stoichiometric ratio. Both of these techniques confirm the unidirectional crystal growth of PdBi$_2$Te$_3$ and its constituent elements to be in near stoichiometric ratio.

We performed room temperature Raman Spectroscopy on as grown PdBi$_2$Te$_3$ crystal and compared it with that of pure Bi$_2$Te$_3$ to understand phonon dynamics, phase purity and changes occurring due to Pd intercalation. Comparison of recorded Raman Spectrum of pure Bi$_2$Te$_3$ and PdBi$_2$Te$_3$ is shown in Fig. 4. Bulk crystal of Bi$_2$Te$_3$ show three vibrational modes viz. $A^1_{1g}$, $E^2_g$ and $A^2_{1g}$. These modes are observed at 60.52, 101.20 and 132.59 cm$^{-1}$ respectively for pure Bi$_2$Te$_3$crystal [22-24]. In case of studied PdBi$_2$Te$_3$ crystal all these three modes are present in Raman Spectra at 60.94, 102.97, 133.47 cm$^{-1}$ which are nearer to that as observed for pure $Bi_2Te_3$ crystal, this shows the phase of grown crystal is similar to that of pure $Bi_2Te_3$. Further, it is observed that these modes are slightly shifted to higher wave number side in comparison to pure $Bi_2Te_3$ and also there is slight change in intensity of peaks. This may be considered due to intercalation of Pd. It seems that Pd intercalation strengthens the bonding forces of $Bi_2Te_3$. Also the width of Raman peaks of studied $PdBi_2Te_3$ crystal is observed to be slightly wider as compared to pure $Bi_2Te_3$ that shows the presence of impurity element in pure $Bi_2Te_3$. Absence of any $A^1_u$ and $E^1_u$ mode between $Eg^2$ and $A^2_{1g}$ shows that Pd intercalation does not break the crystal symmetry in $Bi_2Te_3$ [23]

The Field cooled (FC) and zero field cooled (ZFC) measurements of as grown $PdBi_2Te_3$ crystal under applied of 10 Oe are shown in Fig. 5. Bulk superconductivity is seen in as grown $PdBi_2Te_3$ crystal with $T_c^{onset}$ at below 6K in terms of diamagnetic transition in both zero fields cooled (ZFC) and field cooled (FC) measurements. The transition is sharp and seems near saturating below 2.5K in both FC and ZFC modes. As far as the volume fraction of superconductivity is concerned the same is though low, but is comparable to that as reported earlier [11, 12]. It is suggested earlier that anti-site defects being present in $Bi_2Te_3$ makes it difficult to intercalate Pd in $Bi_2Te_3$ [11]. Another reason being proposed earlier for low volume fraction of topological superconductivity (TS) is the fast degradation of the same with time [4-7].

Plots of magnetization vs applied magnetic field at different temperature viz. 2K, 2.5K, 3K, 3.5K, 4K, 4.5K, 5K, and 6K are shown in Fig. 6, exhibiting bulk type II superconductivity in studied $PdBi_2Te_3$ crystal. These plots show that both the lower and upper critical fields decrease with increasing the temperature and are nearly disappeared below 6K. Figure 7 exhibits the M-H plot of the studied $PdBi_2Te_3$ at 2K. From this plot the values of lower and upper critical field are found to be 180 Oe and 4800 Oe respectively. Here the lower critical field ($H_{c1}$) is defined as the deviation of M-H from linearity and upper critical field ($H_{c2}$) is the closing of the M-H loop coinciding with the irreversibility field. The mean critical field value thus obtained for studied $PdBi_2Te_3$ crystal is $H_c = (H_{c1}*H_{c2})^{1/2}$, which is equal to 929.51 Oe at 2K. Using upper critical field ($H_{c2}$) at 2K, the absolute zero $H_{c2}(0)$ can be calculated with the help of Ginzburg Landau (GL) equation shown in equation 1.

$$H_{c2}(T) = H_{c2}(0) * \left[\frac{1-t^2}{1+t^2}\right] \quad \ldots\ldots \quad (1)$$

Here t is reduced temperature i.e. $t = T/T_c$. The upper critical field is taken at 2K and $H_{c2}(0)$ is calculated by using equation 1, which is found to be 1.02 Tesla. GL parameter κ is calculated by using the relation between $H_c$ and $H_{c2}(0)$ i.e. $H_{c2}(0)=\kappa*2^{1/2}*H_c$, where $H_c$ is mean critical field. Thus calculated κ is found to be 3.65, this value is greater than limit of type I superconductivity which is $1/2^{1/2}$ and thus confirming the type II superconductivity in studied $PdBi_2Te_3$ crystal. Another critical parameter of superconductivity is calculated with the help of formula $H_{c2}(0) = \frac{\varphi_0}{2\pi\xi(0)^2}$, in this formula $\Phi_0$ is a constant, named as flux quanta and its value is 2.0678 x $10^{-15}$Wb. By using above formula ξ(0) i.e. coherence length of the studied $PdBi_2Te_3$ at 0K is found to be 1.72 Å. London penetration depth λ(0) is further calculated by using relation between λ(0) and ξ(0) which is $\kappa = \lambda(0)/\xi(0)$. This relation gives the value of λ(0) to be 6.278 Å. Clearly λ(0) is far greater than ξ(0), confirming again the type II superconductivity in $PdBi_2Te_3$.

Fig. 8 shows expanded M-H plots at different temperature deep right up to superconductivity onset i.e., from 2K to 4.5K. This exercise is done to mark the lower critical field ($H_{c1}$) at different temperatures in superconducting regime. The $H_{c1}$ being marked as deviation from linearity of M-H plots is found to be 180Oe, 150Oe, 122Oe, 107Oe, 90Oe and 68Oe at 2K, 2.5K, 3K, 3.5K, 4K, 4.5K respectively. It is clear that $H_{c1}$ is found to decrease with increasing the temperature. The inset of Fig. 8 depicts the variation of $H_{c1}$ with temperature, showing a monotonic decrement of the same with temperature, which is an obvious behavior of any type II superconductor.

**CONCLUSION**

The short article, reports the single crystal growth and basic superconducting properties of $PdBi_2Te_3$ topological insulator. Phase formation is confirmed by rietveld refinement of powder XRD pattern of the crushed crystal. SEM and EDAX confirmed layered structure and near stoichiometric composition of the studied $PdBi_2Te_3$ topological insulator. Low temperature magnetic measurements showed the presence of type II superconductivity at below 6K, with lower and upper critical fields of 180Oe and 4800Oe respectively at 2K. The GL parameter κ is 3.65, clearly establishing the type II superconductivity in studied $PdBi_2Te_3$ topological insulator.

**ACKNOWLEDGMENTS**

Authors would like to thank Director CSIR-NPL for his keen interest and encouragement. M.M. Sharma would like to thanks CSIR for research fellowship and AcSIR-Ghaziabad for PhD registration.

**FIGURE CAPTIONS**

Fig. 1: Heat Treatment Schematic of the growth for $PdBi_2Te_3$ crystal and inset shows the photograph of a piece of as grown crystal.

Fig. 2: Reitveld fitting of powder x-ray diffraction (PXRD) pattern of as grown $PdBi_2Te_3$ crystal and inset shows the unit cell of the same as drawn by VESTA software.

Fig. 3: Scanning Electron Microscope (SEM) image and EDAX of studied $PdBi_2Te_3$ crystal.

Fig. 4: Room temperature Raman spectrum of $Bi_2Te_3$ and $PdBi_2Te_3$ crystals.

Fig. 5: FC and ZFC magnetization of $PdBi_2Te_3$ single crystal in temperature range of 2K to 6K at an applied field of 10Oe.

Fig. 6: Isothermal magnetization (M-H) of $PdBi_2Te_3$ superconducting single crystal at various temperatures in superconducting regime i.e., between 2K to 6K.

Fig. 7: Isothermal magnetization (M-H) of $PdBi_2Te_3$ superconducting single crystal at 2K, marking the lower ($H_{c1}$) and upper ($H_{c2}$) critical fields.

Fig. 8 Expanded M-H plots of $PdBi_2Te_3$ superconducting single crystal at different temperature deep right up to superconductivity onset i.e., from 2K to 4.5K.

Fig. 1

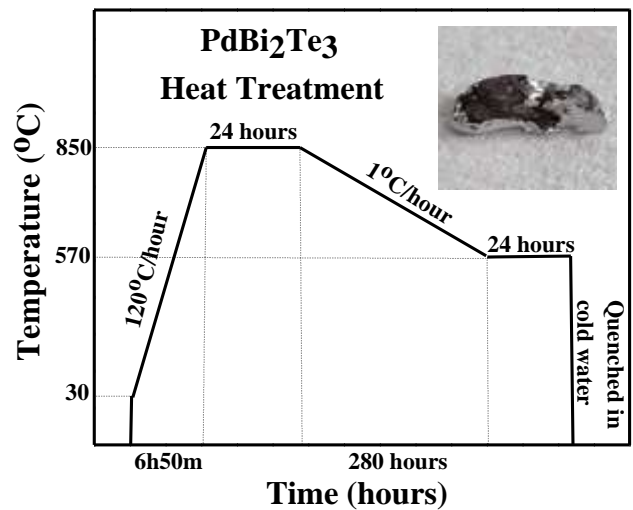

Fig. 2

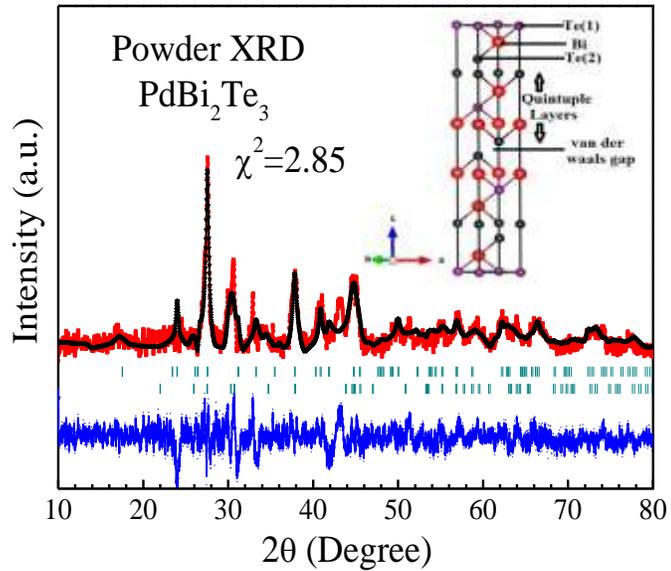

Fig. 3

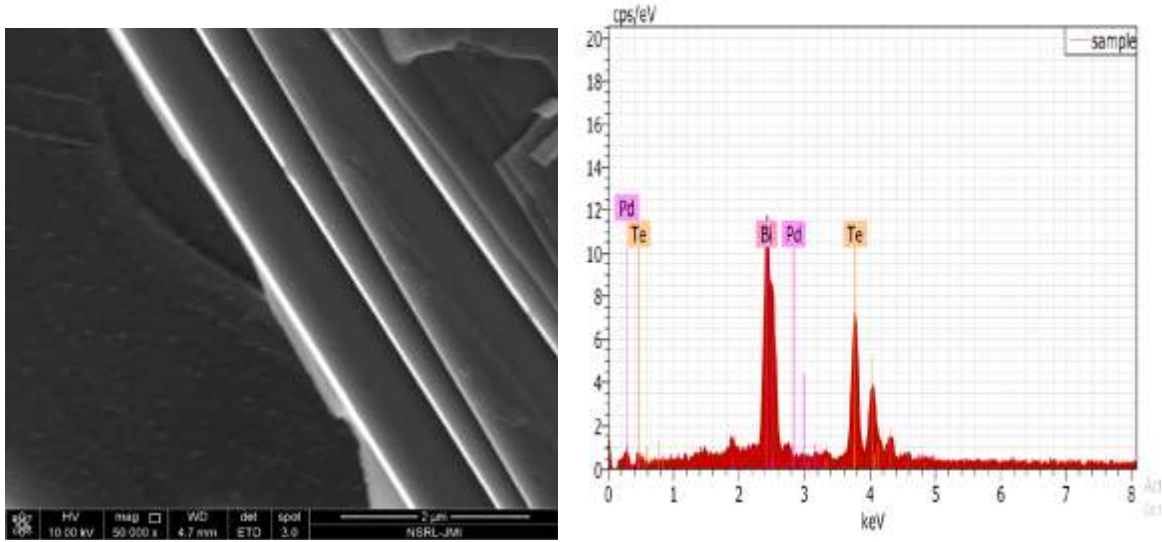

Fig. 4

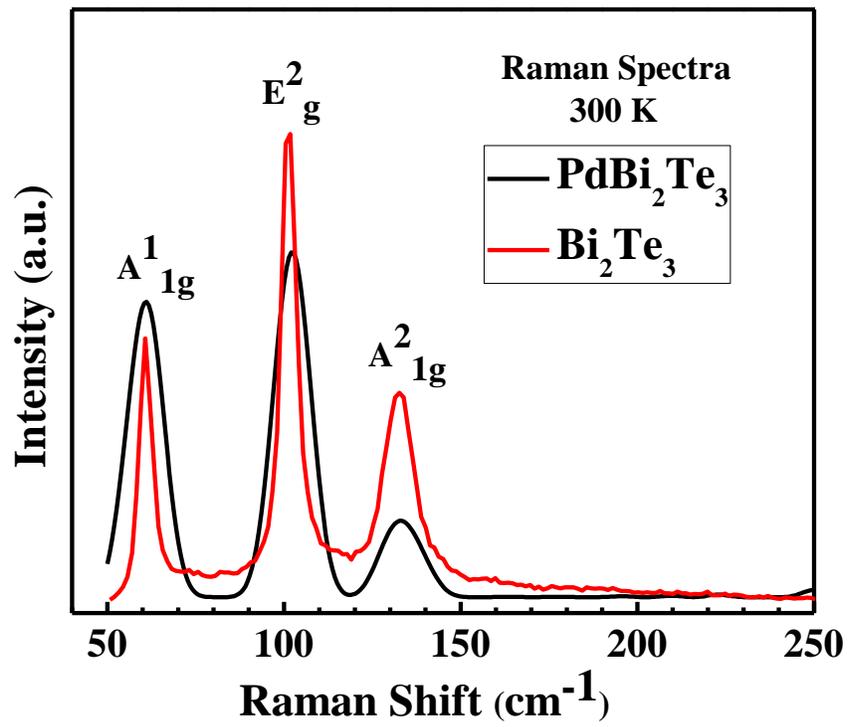

Fig. 5

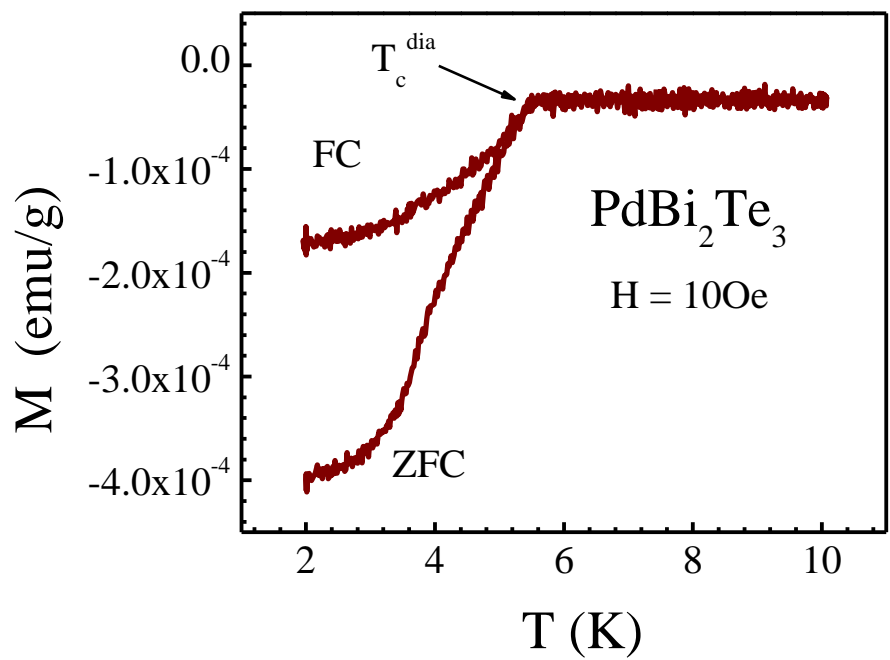

Fig. 6

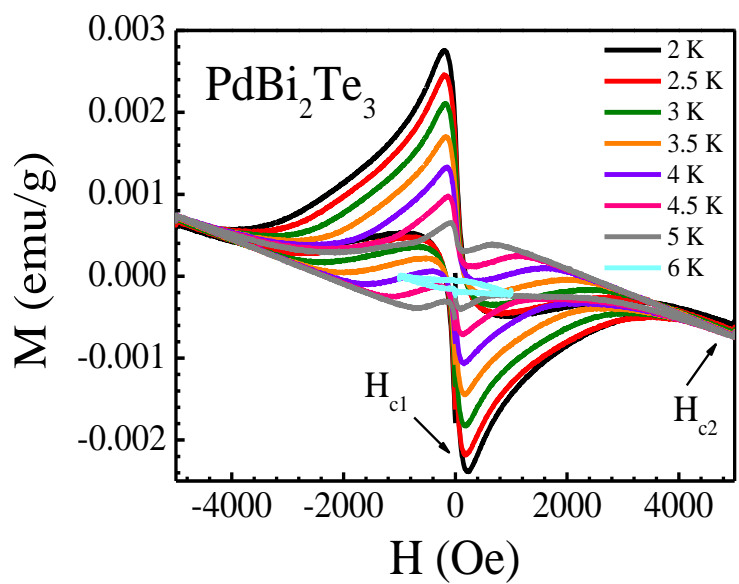

Fig. 7

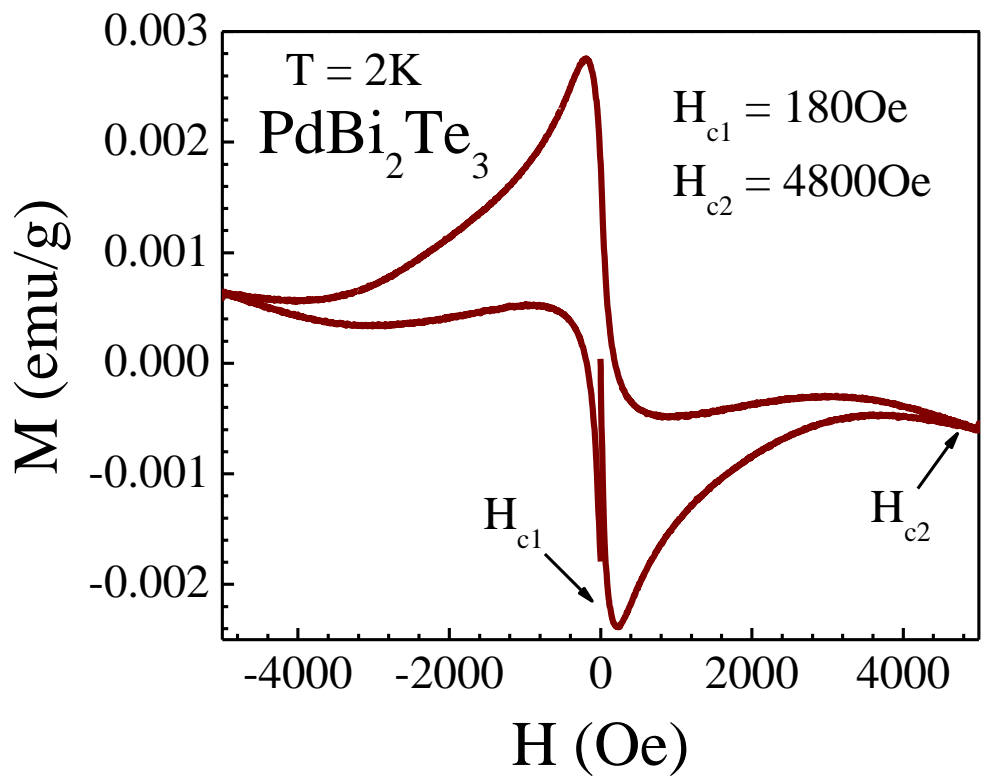

Fig. 8

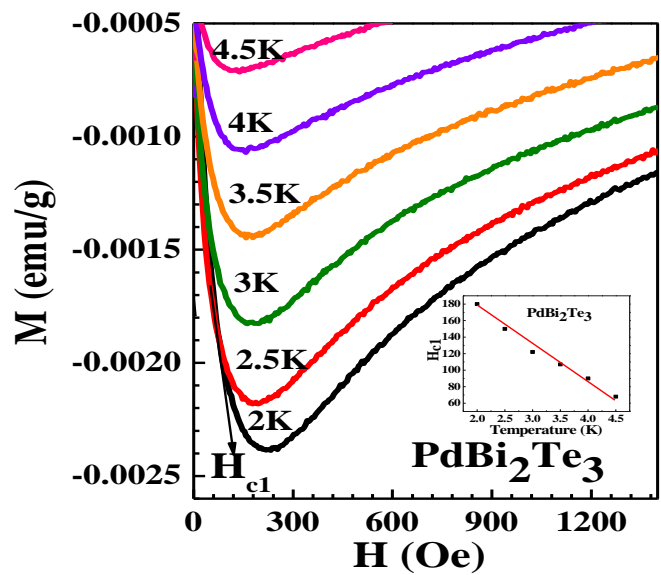